\documentclass[aps,prl,showpacs,twocolumn]{revtex4}
\usepackage{times}
\usepackage{amsmath}
\usepackage{graphicx}
\usepackage{epsfig}
\usepackage{dcolumn}
\usepackage{bm}
\usepackage{color}
\newcommand{\BR}{{\cal B}}
\newcommand{\pp}{\pi^+\pi^-}

\newcommand{\LL}{\ell^+\ell^-}
\newcommand{\EE}{e^+e^-}
\newcommand{\MM}{\mu^+\mu^-}
\newcommand{\GG}{\gamma\gamma}
\newcommand{\psip}{\psi(3686)}

\newcommand{\jpsi}{J/\psi}

\begin{document}
\preprint{\vbox{\hbox{}
                \hbox{Intended for {\it Phys. Rev. Lett.}}
                \hbox{Authors : Liu Zhiqing and Yuan Changzheng}
                \hbox{Committee: Steve Olsen (chair), Du Shuxian and Wu Zhi}
                }}

\title{\quad\\[0.5cm] \boldmath Observation of $\EE\to \eta\jpsi$
at center-of-mass energy $\sqrt{s}=4.009$~GeV}

\author{
M.~Ablikim$^{1}$, M.~N.~Achasov$^{5}$, D.~J.~Ambrose$^{39}$, F.~F.~An$^{1}$, Q.~An$^{40}$, Z.~H.~An$^{1}$, J.~Z.~Bai$^{1}$, Y.~Ban$^{27}$, J.~Becker$^{2}$, J.~V.~Bennett$^{17}$
% ...
\\
\vspace{0.2cm}
(BESIII Collaboration)\\
\vspace{0.2cm} {\it
$^{1}$ Institute of High Energy Physics, Beijing 100049, P. R. China\\
$^{2}$ Bochum Ruhr-University, 44780 Bochum, Germany\\
% ...
$^{5}$ G.I. Budker Institute of Nuclear Physics SB RAS (BINP), Novosibirsk 630090, Russia\\
% ...
$^{17}$ Indiana University, Bloomington, Indiana 47405, USA\\
% ...
$^{27}$ Peking University, Beijing 100871, P. R. China\\
% ...
$^{39}$ University of Rochester, Rochester, New York 14627, USA\\
$^{40}$ University of Science and Technology of China, Hefei 230026, P. R. China\\
% ...
\vspace{0.4cm}
}
\vspace{0.4cm}
}

\date{\today}

\begin{abstract}

Using a $478$~pb$^{-1}$ data sample collected with the BESIII
detector operating at the Beijing Electron Positron Collider storage ring at a center-of-mass
energy of $\sqrt{s}=4.009$~GeV, the production of $\EE\to
\eta\jpsi$ is observed for the first time with a statistical
significance of greater than $10\sigma$. The Born cross section is
measured to be $(32.1\pm 2.8 \pm 1.3)$~pb, where the first error
is statistical and the second systematic. Assuming the $\eta\jpsi$
signal is from a hadronic transition of the $\psi(4040)$, the
fractional transition rate is determined to be $\BR(\psi(4040)\to
\eta\jpsi)=(5.2\pm 0.5\pm 0.2\pm 0.5)\times 10^{-3}$, where the
first, second, and third errors are statistical,
systematic, and the uncertainty from the $\psi(4040)$ resonant parameters,
respectively. The production of $\EE\to \pi^0\jpsi$ is searched
for, but no significant signal is observed, and $\BR(\psi(4040)\to
\pi^0\jpsi) < 2.8 \times 10^{-4}$ is obtained at the 90\%
confidence level.

\end{abstract}


\maketitle

The properties of excited $J^{PC}=1^{--}$ charmonium states above
the $D\bar{D}$ production threshold is of great interest but not well
understood, even decades after their first
observation~\cite{mark3}. The current experimentally
well established structures in the hadronic cross section are the
$\psi(3770)$, $\psi(4040)$, $\psi(4160)$, and
$\psi(4415)$ resonances~\cite{pdg}. Unlike the low-lying vector $c\bar{c}$
states $\jpsi$ and $\psip$, all of these states couple to
open-charm final states with large partial widths, and disfavor
hidden charm decays.

Recently, new vector charmonium-like states, the $Y(4260)$,
the $Y(4360)$ and the $Y(4660)$ have been discovered via their
decays into exclusive $\pp\jpsi$ and $\pp\psip$ final states~\cite{y4260}. The
common properties of these states are relatively narrow widths and
strong couplings to hidden-charm final states. These $Y$-states
cannot be assigned to any of the conventional $c\bar{c}$ $1^{--}$
$\psi$ family states~\cite{potential} in any natural way
and suggest the existence of a non-conventional meson
spectroscopy~\cite{review}.

Hadronic transitions play an important role in understanding
the nature of conventional heavy quarkonium. An excess of
$\eta$ over $\pp$ hidden-bottom transition rates of the $\Upsilon(4S)$~\cite{eta}
has been explained as an admixture of a four-quark state
in the $\Upsilon(4S)$ wave function~\cite{four-quark}. A similar
picture might be expected in the charm sector but, as of yet, there is no
experimental data available for $\eta$ transitions in the high-mass
charmonium and charmoniumlike states, except for evidence of
$\psi(3770)\to\eta\jpsi$ ($3.5\sigma$)~\cite{psi3770} and
$\psi(4160)\to\eta\jpsi$ ($4.0\sigma$)~\cite{sec-cleo}.
Moreover, there are predictions of many new
states in various models trying to explain the conventional and
unconventional states observed in this mass region~\cite{review}.

In this Letter, we report cross section measurements for
$\EE\to\eta\jpsi$ and $\pi^0\jpsi$ at the center-of-mass energy
$\sqrt{s}=(4.009\pm0.001)$~GeV. The analysis is performed with a
478~pb$^{-1}$ data sample collected with the BESIII detector
located at the BEPCII storage ring~\cite{bepc2}. The integrated
luminosity of this data sample was measured using Bhabha events,
with an estimated uncertainty of 1.1\%. In order to control
systematic errors, an accompanying data sample of about seven
million $\psip$ events was accumulated under the same experimental
conditions. In the analysis, the $\jpsi$ is reconstructed through its
decays into lepton pairs ($\EE$ and $\MM$) while $\eta/\pi^0$ is
reconstructed in the $\GG$ final state.

The {\sc geant4}-based Monte Carlo (MC) simulation software, which
includes the geometric description and the detector response, is
used to optimize the event selection criteria, determine the
detection efficiency, and estimate the backgrounds. Signal $\EE\to
\eta\jpsi$ and $\pi^0\jpsi$ MC samples containing 20,000 events
for each channel are generated. Initial state radiation (ISR)
is simulated with {\sc kkmc}~\cite{kkmc}, assuming $\eta\jpsi$ and
$\pi^0\jpsi$ are produced via $\psi(4040)$ decays, and the
$\psi(4040)$ is described by a Breit-Wigner (BW) function with a
constant width. The maximum energies of the ISR photons are
347~MeV and 700~MeV, corresponding to $\eta\jpsi$ and $\pi^0\jpsi$
production thresholds, respectively. For backgrounds studies, MC
samples equivalent to 1~fb$^{-1}$ integrated luminosity are
generated: inclusive $\psi(4040)$ decays, ISR production of
low-mass vector charmonium states, and QED events. The known decay
modes of the charmonium states are generated with {\sc
evtgen}~\cite{evtgen} with branching fractions set to their
world average values~\cite{pdg} and the remaining events are
generated with {\sc lundcharm}~\cite{lund} or {\sc
pythia}~\cite{pythia}.

Charged tracks are reconstructed in the main drift chamber, and the number of
good charged tracks is required to be two with zero net
charge. For each track, the polar angle must satisfy
$|\cos\theta|<0.93$, and the point of closest approach to the
$\EE$ interaction point must be within $\pm 10$~cm in the beam
direction and within $\pm 1$~cm in the plane perpendicular to
the beam direction. A charged track with deposited energy in the
electromagnetic calorimeter less than 0.4~GeV is
identified as a $\mu$ candidate while that with a deposited energy
over momentum ($E/p$) ratio larger than 0.8 is identified as an
electron candidate. Both of the two charged tracks are required to
be either identified as muons or as electrons.

Showers identified as photon candidates must satisfy fiducial and
shower-quality requirements. The minimum energy is 25~MeV for electromagnetic calorimeter
barrel showers ($|\cos\theta|<0.8$) and 50~MeV for end-cap showers
($0.86<|\cos\theta|<0.92$). To eliminate showers produced by
charged particles, a photon must be separated by at least 20
degrees from any charged track. Final state radiation (FSR) and
bremsstrahlung energy loss of leptons are corrected by adding
the momentum of photons detected within a 5 degree cone around
the lepton momentum direction. The number of good photon
candidates is required to be two (the efficiency is
over 95\%), and the recoil mass of the two
photons $M_{\rm recoil}(\GG)=\sqrt{(P_{\rm CM}-P1-P2)^2}\in
[2.9, 3.4]~{\rm GeV}/c^2$ is required to select good $\jpsi$
candidates. Here $P_{\rm CM}$ is the four-momentum of the initial
states, and $P1$, $P2$ are the four-momenta of the two photons.

The lepton pair and the two photons are subject to a
four-constraint (4C) kinematic fit to improve the momentum
resolution and reduce the background. The chi-square ($\chi^2$) of
the kinematic fit is required to be less than 40. In order to
reject radiative Bhabha and radiative dimuon ($\gamma\EE/\gamma\MM$) backgrounds associated
with an energetic radiative photon ($\gamma_H$) and a low
energy fake photon, the invariant mass $M(\gamma_H\LL)$ is
determined from a three-constraint (3C) kinematic fit in which the
energy of the low energy photon is allowed to float. Since
the fake photon does not contribute in the 3C-fit, the
$M(\gamma_H\LL)$ mass distribution is not distorted by the
photon energy threshold cutoff,
and backgrounds are clearly separated from signal.
The requirement $M(\gamma_H\LL)<3.93$~GeV/c$^2$ removes
over 50\% of radiative Bhabha and radiative dimuon
background events with an efficiency greater than 99\% for
$\eta\jpsi$ and 89\% for $\pi^0\jpsi$.

\begin{figure*}[htbp]
\begin{center}
\includegraphics[height=2.3in]{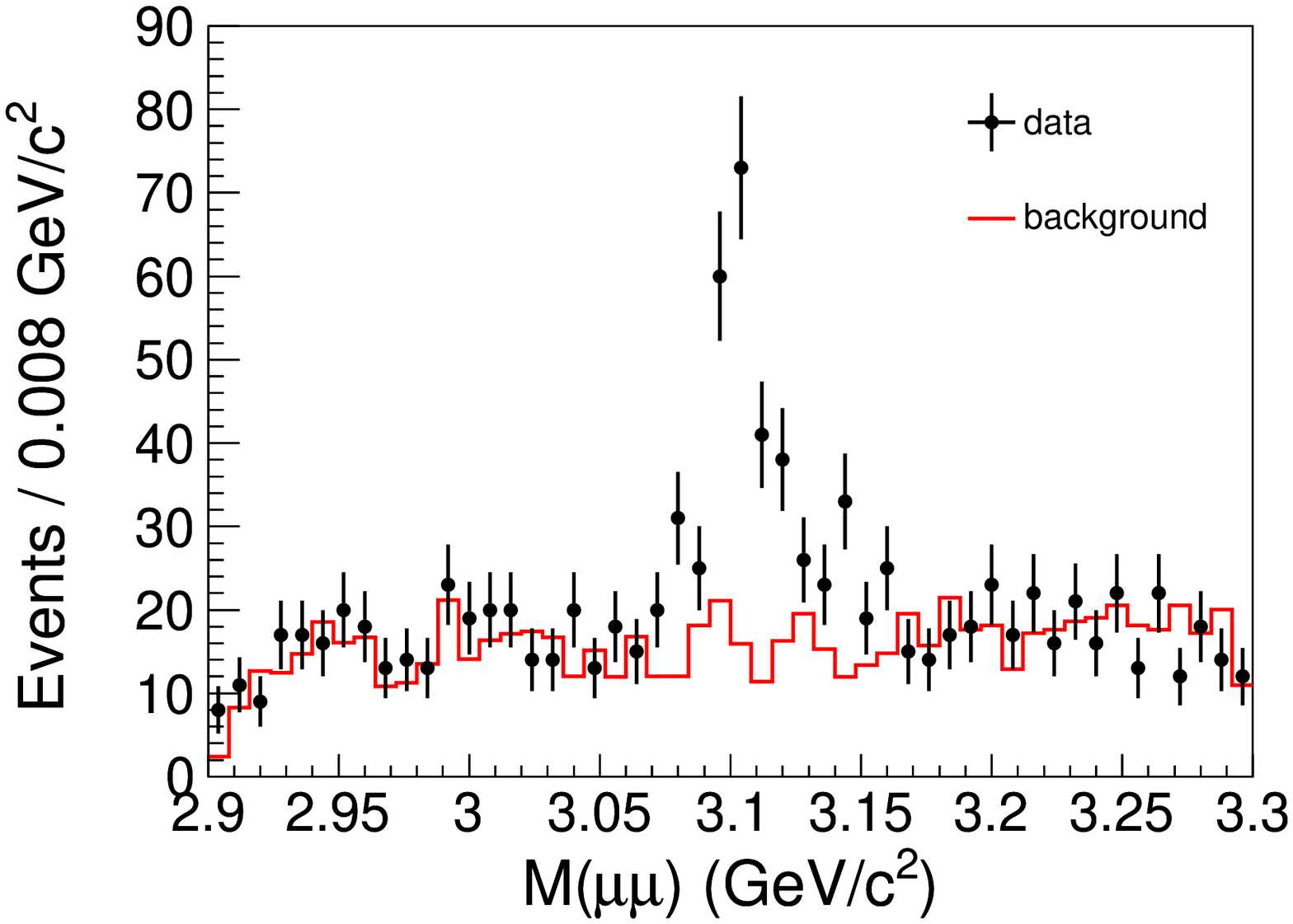}
\includegraphics[height=2.3in]{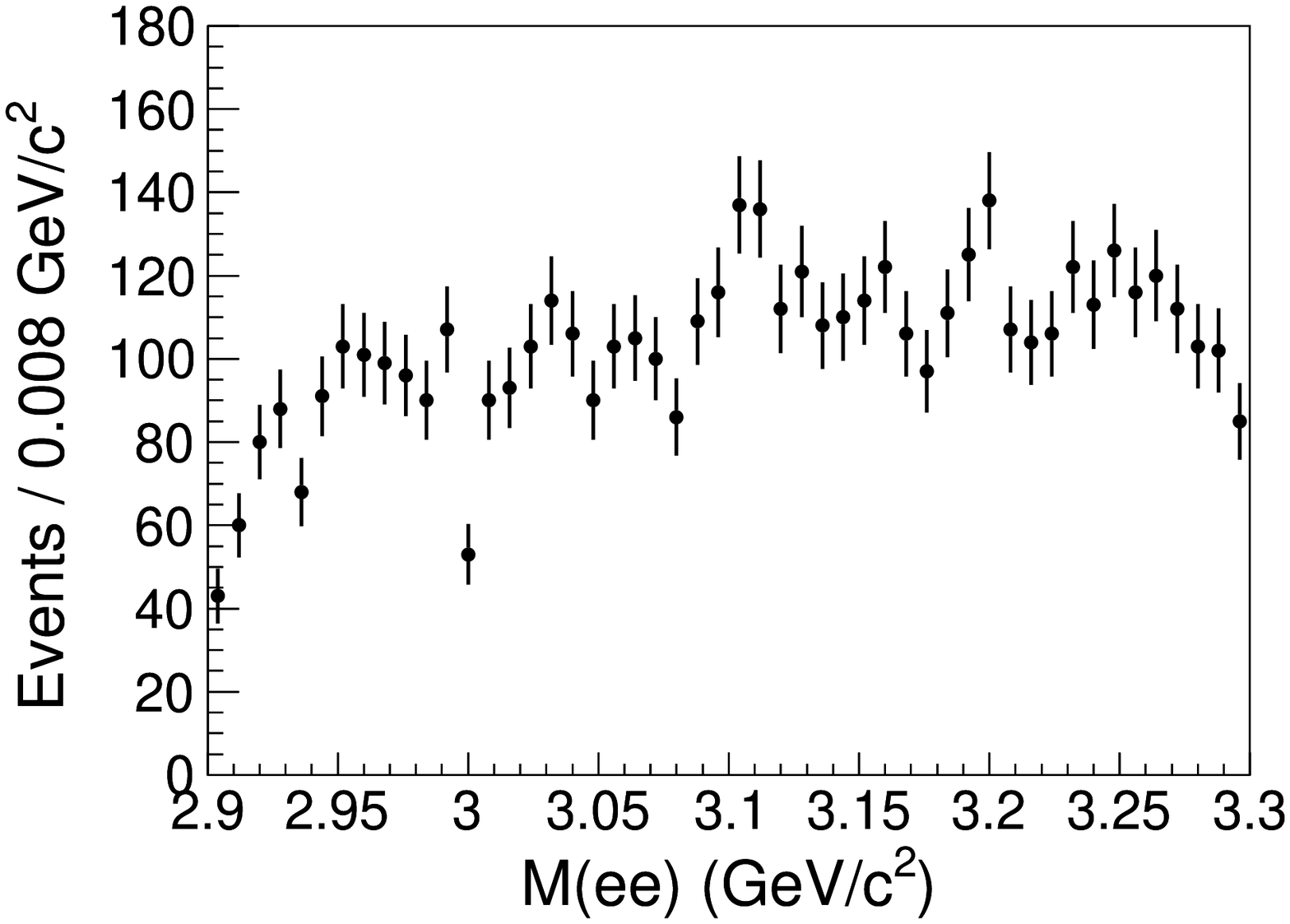}
\caption{(Left panel) $M(\MM)$ and (right panel) $M(\EE)$
invariant mass distributions. Dots with error bars are data and
the open histogram in the left panel shows inclusive-MC-estimated
background events.} \label{m2l}
\end{center}
\end{figure*}

After imposing all of these selection criteria, the invariant
mass distribution of lepton pairs is shown in Fig.~\ref{m2l}.
A clear $\jpsi$ signal is observed in the $\MM$ mode while
indications of a peak
around 3.1~GeV/c$^2$ also exist in the $\EE$ mode. The remaining
dominant backgrounds are surviving radiative dimuon events in
$\MM$ and radiative Bhabha events in $\EE$; these
contribute flat components in the $M(\LL)$ distributions with
no associated peaks in the $M(\GG)$ invariant mass distribution. The high
background level in the $\EE$ mode is due to the huge background from
the Bhabha process. Other possible background sources include $\EE\to
\pi^0\pi^0\jpsi$, $\pp\pi^0/ \pp\eta$, and $\gamma\chi_{cJ}(1P)/\gamma\chi_{cJ}(2P)$.
The $\pi^0\pi^0\jpsi$ background is estimated by MC simulation to be at
the 4.5~pb level and, thus, negligibly small~\cite{sec-cleo}. Potential
$\gamma\chi_{cJ}(1P)$ and $\gamma\chi_{cJ}(2P)$ radiative transition backgrounds are
estimated using the selected data sample; no significant signal
is found for either $\chi_{cJ}(1P)$ or $\chi_{cJ}(2P)$ in $M(\gamma\jpsi)$ mass
distribution. The $\pp\pi^0$ and $\pp\eta$ backgrounds are estimated
using $\jpsi$ sideband events. The ISR-produced vector
charmonium backgrounds, including $\gamma_{\rm ISR}\jpsi$,
$\gamma_{\rm ISR}\psip$ and $\gamma_{\rm ISR}\psi(3770)$, are
estimated by means of an inclusive MC sample and only 3.3 events in the $\MM$
mode and 3.1 events in the $\EE$ mode are found (normalized to data
luminosity). As they would peak at neither the $\eta$ nor the $\pi^0$
signal region, they are neglected in the analysis.

\begin{figure*}[htbp]
\begin{center}
\includegraphics[height=2.3in]{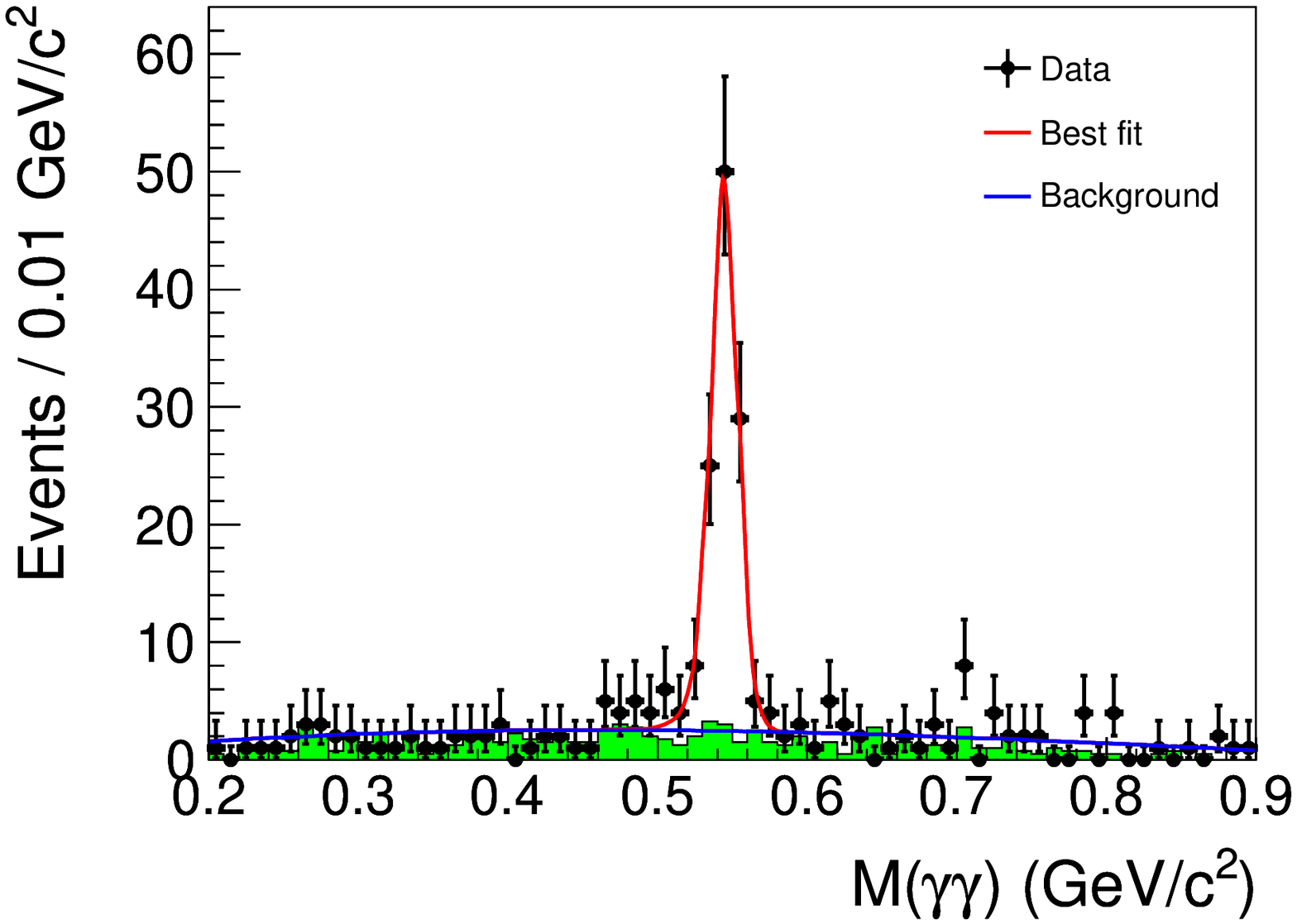}
\includegraphics[height=2.3in]{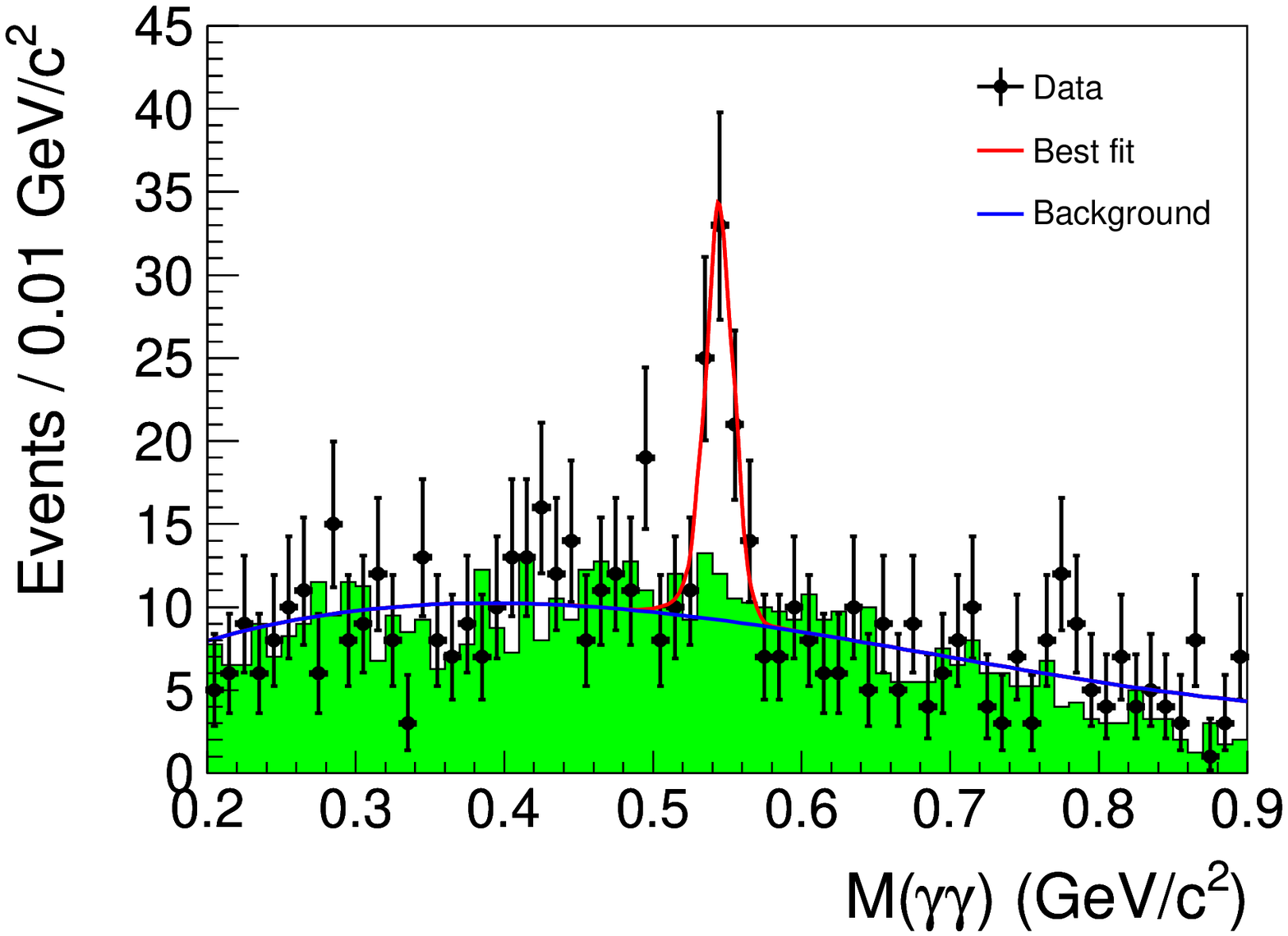}
\caption{ Distributions of $M(\GG)$ between 0.2~GeV/c$^2$ and 0.9~GeV/c$^2$
for $\jpsi\to \MM$ (left panel) and for
$\jpsi\to \EE$ (right panel). Dots with error
bars are data in $\jpsi$ mass signal region, and the green shaded
histograms are from normalized $\jpsi$ mass sidebands. The curves
show the total fit and the background term.} \label{fit-mgg}
\end{center}
\end{figure*}

\begin{figure*}[htbp]
\begin{center}
\includegraphics[height=2.3in]{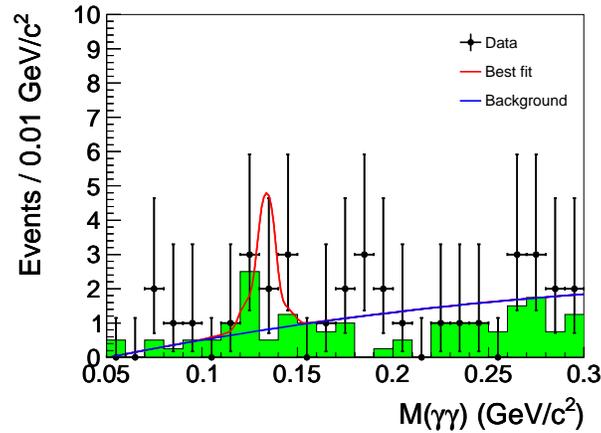}
\caption{ Distribution of $M(\GG)$ below 0.3~GeV/c$^2$
for $\jpsi\to \MM$. Dots with error
bars are data in $\jpsi$ mass signal region, and the green shaded
histogram is from normalized $\jpsi$ mass sideband. The curves
show the total fit and the background term.}
\label{fit-pi0}
\end{center}
\end{figure*}

The resolution of the invariant mass of the lepton pairs is determined
to be 14~MeV/c$^2$ by MC simulation, and is in good agreement with
events in the $\psip$ data sample. The mass window of the $\jpsi$
signal is defined as 3.075~GeV/c$^2$ $<M(\LL)<$ 3.125~GeV/c$^2$, and
the sidebands are defined as 2.95~GeV/c$^2$ $<M(\LL)<$ 3.05~GeV/c$^2$
or 3.15~GeV/c$^2$ $<M(\LL)<$ 3.25~GeV/c$^2$, which is four times as
wide as the signal region. Figure~\ref{fit-mgg} shows the $M(\GG)$
invariant mass distributions for events in the $\jpsi\to \MM$ and
$\jpsi\to \EE$ signal regions. A significant $\eta$ signal is observed
in both modes. In the $M(\GG)$ distribution for $\jpsi$ mass-sideband
events, there are backgrounds that peak in the $\pi^0$ signal region in $\jpsi\to\MM$ that
originate from $\EE\to \pp\pi^0$. In order to suppress $\EE\to\pp\pi^0$
backgrounds, at least one charged track is required to have a muon counter hit
depth larger than 30~cm for the $\pi^0\jpsi$ signal search. The
efficiency for this requirement is 87.9\% for signal while about
74\% $\EE\to \pp\pi^0$ background events are
rejected. Figure~\ref{fit-pi0} shows the $M(\GG)$ invariant mass
distribution below 0.3~GeV/c$^2$ for $\jpsi\to \MM$. No
significant $\pi^0$ signal is observed. We do not analyze $\pi^0\jpsi$
production in $\jpsi\to \EE$ due to the huge background from Bhabha
events. The final selection efficiencies are 38.0\% in $\MM$
and 26.9\% in $\EE$ for $\eta\jpsi$, and 31.1\% in $\MM$
for $\pi^0\jpsi$, according to MC simulation.

The $M(\GG)$ invariant mass distributions are fitted using an unbinned
maximum likelihood method for $M(\GG)<0.9$~GeV/c$^2$ in both
modes. The probability density function (pdf) for the $\eta$/$\pi^0$
signal in $\jpsi\to\MM$ is taken from MC simulation, while in
$\jpsi\to\EE$, only the $\eta$ pdf from MC simulation is used.  To
account for resolution differences between data and the MC simulation,
three Gaussian functions are convolved with the $\eta$ and the $\pi^0$
signal pdfs. For the $\eta$ signal, the standard deviation of these
Gaussians are free while for $\pi^0$ signal, it is fixed to $(2.4\pm
0.9)$~MeV/c$^2$, which is determined from a $\psip\to \pi^0\jpsi$
control sample. Background shapes are described by a third-order
polynomial. Figure~\ref{fit-mgg} shows the fit results for the $\eta$
signal and the background contributions for $\jpsi\to \MM$ and
$\jpsi\to \EE$. The fits yield $N^{\rm fit}_{\MM}(\eta) = 111.4\pm
11.0$, and $N^{\rm fit}_{\EE}(\eta) = 61.4\pm 10.5$. The standard
deviation of the smearing Gaussian convolved with the $\eta$ signal is
$(3.7\pm 1.0)$~MeV/c$^2$ in $\MM$  and $(3.7\pm
1.9)$~MeV/c$^2$ in  $\EE$. Good agreement is observed between the
two modes, and these values are consistent with values from the
$\psip\to \eta\jpsi$ control sample ($3.4\pm0.6$~MeV/c$^2$ in $\MM$
and $4.6\pm0.6$~MeV/c$^2$ in $\EE$). The goodness of fit is
estimated by using a $\chi^2$ test method with the data distributions
regrouped to ensure that each bin contains more than 10 events.  The
test gives $\chi^2/n.d.f$=14.1/14=1.1 for $\MM$ and
$\chi^2/n.d.f$=42.9/43=1.0 for $\EE$.  Figure~\ref{fit-pi0} shows
the fit result for the $\pi^0$ signal and the background contribution
for $\jpsi\to \MM$.  Since the $\pi^0$ signal is not
significant, we determine an upper limit for the $\pi^0$ signal yield
of $N^{{\rm up}}(\pi^0)<11.7$ at the 90\% confidence level. The
$\EE\to\pp\pi^0$ backgrounds are estimated by fitting the $M(\GG)$ distribution
of the $\jpsi$ mass sideband events. The signal pdf for the $\pi^0$ is
a Gaussian function and that for the background is a third-order
polynomial.  The fit yields $N^{\rm bkg}_{\MM}(\pi^0)=2.8\pm 1.1$
after normalization. The statistical significances of the $\eta$ and
$\pi^0$ signals are examined by means of the difference in
log-likelihood value with or without signal in the fit and the change
of the number of degrees of freedom ($\Delta$ndf). For the $\eta$
signal, the statistical significance is larger than $10\sigma$ while
that for the $\pi^0$ signal is only $1.1\sigma$.

The Born-order cross section is determined from the relation
\begin{equation}
 \sigma^{B}=\frac{N^{\rm fit}-N^{\rm bkg}}
 {\mathcal{L}_{\rm int}(1+\delta) \epsilon \BR},
\end{equation}
where $N^{\rm fit}$ and $N^{\rm bkg}$ are the number of signal
events from the fit and the number of background events,
respectively; $\mathcal{L}_{\rm int}$ is integrated luminosity;
$\epsilon$ is selection efficiency; $\BR$ is branching fraction of
intermediate states decay; and ($1+\delta$) is the radiative
correction factor, which is 0.757 according to QED
calculation~\cite{rad}.

For the $\EE\to\eta\jpsi$ cross section,
we obtain $\sigma^B = 34.8\pm 3.5$~pb for the $\MM$
mode, and $\sigma^B = 27.1\pm 4.7$~pb for the $\EE$ mode. Since the results
from the two modes agree with each other, we quote a combined cross
section result:
\begin{equation}
 \sigma^B(\EE \to \eta\jpsi) = 32.1\pm 2.8~{\rm pb}.
\end{equation}
Here the errors are statistical only.

Systematic errors mainly come from the luminosity measurement,
detection efficiency, background estimation and branching fractions
of intermediate states decays. All the contributions
are summarized in Table~\ref{total-err}.

\begin{table}
\caption{Summary of the systematic errors (\%) in the cross
section measurement.} \label{total-err}
\begin{center}
\begin{tabular}{cccc}
  \hline\hline
  Source & $\eta\MM$ & $\eta\EE$ & $\pi^0\MM$\\
  \hline
  Luminosity & 1.1 & 1.1 & 1.1\\
  Tracking & 2 & - & 2\\
  Photon detection  & 2 & 2 & 2\\
  Lepton resolution & 1.6 & 2.4 & 1.6\\
  Kinematic fit & 1.9 & 1.9 & 1.9\\
  Background shape & 1.5 & 3.0 & 9.4\\
  Fit function & - & - & 3.9\\
  $\psi(4040)$ parameters & 2.0 & 3.3 & 4.0\\
  Branching fractions & 1.2 & 1.2 & 1.0\\
  Others & 1.0 & 1.0 & 1.0 \\\hline
  Total & 5.0 & 6.1 & 11.8\\
  \hline\hline
\end{tabular}
\end{center}
\end{table}

The uncertainty from luminosity measurement is estimated to be
1.1\% using Bhabha events. The muon tracking efficiency is
estimated to be 1\% for each track. Since the luminosity is measured
using Bhabha events, the tracking efficiency of electron pairs
cancels. The photon
detection efficiency is also estimated to be 1\% for each photon.
The uncertainties associated with the lepton pair invariant mass
resolutions and the kinematic fits are estimated using the $\psip\to
\eta\jpsi$ control sample. It is obtained from the $\psip$
data sample by imposing the selection criteria
described above, and requiring
$M(\gamma_H\jpsi)<3.49$~GeV/c$^2$ to reject $\chi_{c1}$ and $\chi_{c2}$ events.
This gives a low-background $\psip\to \eta\jpsi$ events with a
purity of 98.5\%. The efficiency difference between data and MC simulation
for the $\jpsi$ invariant mass window is 1.6\% in the $\MM$ mode
and 2.4\% in the $\EE$ mode. They are taken as systematic errors
due to lepton-pair invariant mass resolution. For the kinematic fit,
the efficiency difference between data and MC simulation is 1.9\%
in both modes.

Uncertainties due to the choice of background shape are estimated by varying
the background function from a 3rd-order polynomial to a 2nd-order and
a 4th-order polynomial in the fit, and these changes yield a 1.5\% difference
in $\MM$ and a 3.0\% difference in $\EE$ in the number of $\eta$
signal events. The $\EE\to\pp\pi^0$ backgrounds subtraction gives a 9.4\% difference
in $\MM$ in the number of $\pi^0$ signal events. The uncertainty due
to the fit function is estimated by changing the smearing Gaussian parameter
by one standard deviation in the $\pi^0$ signal pdf, which gives 3.9\% difference
in the number of $\pi^0$ signal events. Uncertainties in the
$\psi(4040)$ resonance parameters and possible distortions of the
$\psi(4040)$ line shape due to interference effects with the nearby
$\psi(4160)$ resonance introduce uncertainties in
the radiative correction factor and the efficiency. Changing the
Breit-Wigner parameters (mass and width) by one standard deviation
according to PDG values~\cite{pdg}, or using a coherent shape with
the $\psi(4160)$ resonance~\cite{fpcp} result in variations in
$(1+\delta)\times\epsilon$ of 2.0\% in $\MM$  and 3.3\% in
 $\EE$  for the $\eta\jpsi$ measurement, and, 4.0\% in
$\MM$ for $\pi^0\jpsi$ measurement. The PDG uncertainty in
$\BR(\jpsi\to\LL)$ is 1\% and $\BR(\eta\to\GG)$
is 0.5\%~\cite{pdg}. Other sources of systematic
error, including fake photon simulation and the final-state radiation
simulation, are estimated to be 1.0\% in total.

Assuming all the sources are independent, the total systematic
errors on the $\eta\jpsi$ cross section measurement is determined to be
5.0\% for $\MM$ and 6.1\% for $\EE$. Considering the
common and uncommon errors for these two modes, the combined systematic error
on the $\eta\jpsi$ cross section measurement is 4.0\%. The total
systematic error is 11.8\% in $\MM$ for the $\pi^0\jpsi$ cross
section measurement by summing up all the errors in quadrature.

Since the significance of the $\pi^0\jpsi$ signal is low, an upper limit
on the $\pi^0\jpsi$ production cross section is set at
$\sigma^{B}(\EE\to \pi^0\jpsi)<1.6$~pb at the 90\% confidence level,
where $\EE\to\pp\pi^0$ backgrounds have been subtracted and the efficiency
is lowered by a factor of ($1-\sigma_{sys}$).

If we assume the observed $\eta\jpsi$ and $\pi^0\jpsi$ are
completely from $\psi(4040)$ decays and use the total cross
section of $\psi(4040)$ at $\sqrt{s}=4.009$~GeV [$(6.2\pm 0.6)$~nb]
calculated with the PDG resonance parameters~\cite{pdg} as
input, we determine the fractional transition rate $\BR(\psi(4040)\to \eta\jpsi) =
(5.2\pm 0.5\pm 0.2\pm 0.5)\times 10^{-3}$, where the first,
second, and third errors are statistical, systematic, and
uncertainty from $\psi(4040)$ resonant parameters, respectively.
In addition, we obtain an upper limit on $\BR(\psi(4040)\to \pi^0\jpsi) <
2.8 \times 10^{-4}$ at the 90\% confidence level.

In summary, we observe for the first time $\EE\to \eta\jpsi$
production at $\sqrt{s}=4.009$~GeV with a statistical significance
greater than 10$\sigma$. The Born cross section is measured to be
$(32.1\pm 2.8 \pm 1.3)$~pb, where the first error is statistical
and second systematic. We do not observe a significant $\EE\to
\pi^0\jpsi$ signal, and the Born cross section is found to be less
than 1.6~pb at the 90\% confidence level. These measurements do not
contradict the upper limits set by CLEO experiment~\cite{sec-cleo}.
The $\eta\jpsi$ cross section measurement is within the range of the
theoretical calculation and the $\pi^0\jpsi$ upper limit
does not exclude the prediction~\cite{theory}.
A transition rate of $5\times 10^{-3}$
level is measured for $\psi(4040)\to \eta\jpsi$, corresponding to
a partial decay width at the 400~keV level, which is much larger than
that for $\psi(3770)\to \eta\jpsi$~\cite{psi3770} and is more than two times of
that for $\psi(4040)\to \pp\jpsi$~\cite{sec-cleo}.

The BESIII collaboration thanks the staff of BEPCII and the computing
center for their hard efforts. This work is supported in part by the
Ministry of Science and Technology of China under Contract No. 2009CB825200;
National Natural Science Foundation of China (NSFC) under Contracts
Nos. 10625524, 10821063, 10825524, 10835001, 10935007, 11125525, 11235011, 11205163; Joint
Funds of the National Natural Science Foundation of China under Contracts
Nos. 11079008, 11079027, 11179007; the Chinese Academy of Sciences (CAS) Large-Scale
Scientific Facility Program; CAS under Contracts Nos. KJCX2-YW-N29,
KJCX2-YW-N45; 100 Talents Program of CAS; Istituto Nazionale di Fisica Nucleare,
Italy; Ministry of Development of Turkey under Contract No. DPT2006K-120470;
U. S. Department of Energy under Contracts Nos. DE-FG02-04ER41291,
DE-FG02-91ER40682, DE-FG02-94ER40823; U.S. National Science Foundation;
University of Groningen (RuG) and the Helmholtzzentrum fuer Schwerionenforschung
GmbH (GSI), Darmstadt; WCU Program of National Research Foundation of
Korea under Contract No. R32-2008-000-10155-0.


\begin{thebibliography}{**}

\bibitem{mark3} J. Siegrist {\em et al.}, Phys. Rev. Lett.
{\bf 36}, 700 (1976).

\bibitem{pdg} J. Beringer {\em et al.} (Particle Data Group),
Phys. Rev. D {\bf 86}, 010001 (2012).

\bibitem{y4260} B. Aubert {\em et al.} ({\em BABAR} Collaboration),
Phys. Rev. Lett. {\bf 95}, 142001 (2005); C. Z. Yuan {\em et al.}
(Belle Collaboration), Phys. Rev. Lett. {\bf 99}, 182004 (2007);
X. L. Wang {\em et al.} (Belle Collaboration), Phys. Rev. Lett.
{\bf 99}, 142002 (2007).

\bibitem{potential} E. Eichten {\em et al.}, Phys. Rev. D {\bf 17},
3090 (1978); {\bf 21}, 203 (1980); T. Barnes, S. Godfrey, and E.
S. Swanson, Phys. Rev. D {\bf 72}, 054026 (2005).

\bibitem{review} For a recent review, see
  N.~Brambilla {\em et al.},
  Eur.\ Phys.\ J.\ C {\bf 71}, 1534 (2011).

\bibitem{eta} B. Aubert {\em et al.} ({\em BABAR} Collaboration),
Phys. Rev. D {\bf 78}, 112002 (2008).

\bibitem{four-quark} M. B. Voloshin,
Mod. Phys. Lett. A {\bf 26}, 773 (2011).

\bibitem{psi3770} N. E. Adam {\em et al.} (CLEO Collaboration),
Phys. Rev. Lett. {\bf 96}, 082004 (2006).

\bibitem{sec-cleo} T. E. Coan {\em et al.} (CLEO Collaboration),
Phys. Rev. Lett. {\bf 96}, 162003 (2006).

\bibitem{bepc2} M. Ablikim {\em et al.} (BESIII Collaboration),
Nucl. Instrum. Methods Phys. Res., Sect. A {\bf 614}, 345 (2010).

\bibitem{kkmc} S. Jadach, B. F. L. Ward, and Z. Was,
Comput. Phys. Commun. {\bf 130}, 260 (2000); Phys. Rev. D {\bf
63}, 113009 (2001).

\bibitem{evtgen} R. G. Ping {\em et al.}, Chinese Phys. C {\bf 32}, 599 (2008).

\bibitem{lund} Wang Ping, Ma Yan-Yun, Qin Xiu-Bo, Zhang Zhe, Cao Xing-Zhong, Yu Run-Sheng and Wang Bao-Yi, Chinese Phys. C {\bf 32},
243 (2008).

\bibitem{pythia} http://home.thep.lu.se/$\sim$torbjorn/Pythia.html

\bibitem{rad} E. A. Kuraev and V. S. Fadin, Yad. Fiz. {\bf 41},
733 (1985) [Sov. J. Nucl. Phys. 41, 466 (1985)].

\bibitem{fpcp} X. L. Wang for the Belle Collaboration, talk at the
FPCP meeting [http://hepg-work.ustc.edu.cn/fpcp2012].

\bibitem{theory} Q. Wang, G. Li, X. H. Liu and Q. Zhao, arXiv:1206.4511.

\end{thebibliography}
\end{document}